\documentclass[]{jfm}

\usepackage{graphicx}
\usepackage{newtxtext}
\usepackage{newtxmath}
\usepackage{natbib}
\usepackage{hyperref}
\hypersetup{
    colorlinks = true,
    urlcolor   = blue,
    citecolor  = black,
}

\newcommand{\RomanNumeralCaps}[1]

\title{Dynamics of a microroller under confinement}

\author{
Han Gao\aff{1,3}, 
Nan Xie\aff{2}, 
Zaiyi Shen\aff{4}, 
Xiaoping Hu\aff{2}, 
Shiyuan Hu\aff{2}\corresp{\email{shiyuanhu@buaa.edu.cn}}
\and Ye Xu\aff{1,3} \corresp{\email{ye.xu@buaa.edu.cn}}
}

\affiliation{\aff{1}Hangzhou International Innovation Institute, Beihang University, Hangzhou 311115, China
\aff{2}School of Physics, Beihang University, Beijing 102206, China
\aff{3}School of Mechanical Engineering and Automation, Beihang University, 
Beijing 102206, China
\aff{4}State Key Laboratory for Turbulence and Complex Systems, School of Mechanics and Engineering Science,
Peking University, Beijing 100871, China}

\begin{document}
\maketitle

\begin{abstract}
Rotating particles can translate when placed near a surface, forming microrollers with a wide range of biomedical and microfluidic applications. In this work, we investigate the dynamics of microrollers in confined microchannels with different geometries by combining experiments, numerical simulations, and scaling analysis. In constricted channels, we find that the translational velocity of a microroller decreases as it approaches the constricted region. In both rectangular and cylindrical channels, velocity reversal occurs as the characteristic channel width decreases. Using the force-free condition for free translation, we develop a systematic scaling framework that can be generalized to different channel geometries. The scaling analysis yields functional dependences of the translational velocity on the degree of confinement, which agree well with both experiments and simulations. Importantly, we demonstrate that the viscous stress generated by the far-field rotlet flow governs the observed velocity reduction and reversal, while the translational resistance resulting from the near-field shear flow suppresses translation under tight confinement. The distinct roles of these flow components revealed by our analysis may provide practical guidance for controlling microroller dynamics in confined fluid environments.
\end{abstract}

\begin{keywords}

\end{keywords}


\section{Introduction}\label{sec:introduction}

In viscous fluids at the microscale, propulsion plays a crucial role not only in biological systems~\citep{Lauga2009} but also in numerous biomedical and microfluidic applications, including fluid pumping~\citep{Gu2020,Hu2023}, targeted delivery, and cell manipulation~\citep{Sitti2015}. To break time reversal symmetry and generate net propulsion at low Reynolds number (\Rey)~\citep{Purcell77}, various artificial swimmers have been developed, including flagellar or ciliary swimmers~\citep{Dreyfus2005}, three-sphere swimmers~\citep{Najafi2004}, and phoretic active particles~\citep{Moran2017}. In recent years, rotating magnetic particles driven by external time-dependent magnetic fields have attracted considerable attention~\citep[see e.g.,][]{Xie2019,Zhou21}. When placed near a surface, hydrodynamic interactions with the boundary convert particle rotation into translation~\citep{Goldman1967}, producing microrollers whose translational velocity depends sensitively on proximity to the surface. 

The dynamics of an individual microroller is influenced by factors such as its shape~\citep{Tierno2008,Bozuyuk2021}, magnetization properties~\citep{Tierno2007,Janssen2009}, and fluid viscoelasticity~\citep{Gao25,He2026}. In particular, a critical driving frequency exists, beyond which the microroller no longer rotates synchronously with the magnetic field~\citep{Cebers2006,Mahoney2014}. These dependencies may be exploited to use microrollers as microsensors for probing relevant physical properties~\citep{McNaughton2007}. The motion of a microroller generates a disturbance flow field, which can be approximated in the far field by a rotlet singularity located near the boundary~\citep{Blake1974}. Beyond individual motion, a variety of collective behaviors emerge in microroller swarms mediated by such hydrodynamic interactions~\citep{Martinez2015,Driscoll2017,Yu2019}. 

Since the directed motion of microrollers is assisted by bounding surfaces, microrollers and their assemblies are well suited to operate in confined fluid environments, such as in microfluidic channels and biological tubular structures~\citep{Zhou2024}. It has been shown that confinement can significantly affect the dynamics and flow fields of both swimming microorganisms~\citep{Jeanneret2019,Vizsnyiczai2020,Da2025} and synthetic swimmers~\citep{Takagi2014}. Owing to the long-range nature of hydrodynamic interactions at low \Rey, microroller dynamics is also expected to be strongly influenced by confinement. When confined between two parallel walls, simulations show that the velocity of a microroller decreases as the wall separation decreases~\citep{Fang2020}. Below a critical wall separation, the translational velocity reverses direction, moving opposite to the rolling direction. Experiments further confirm the reduction in velocity with increasing confinement and demonstrate that a microroller cannot enter a locally confined region when the confinement becomes sufficiently strong~\citep{Bozuyuk2022}. Velocity reversal has also been observed in more complex geometries, including cylindrical and rectangular channels~\citep{Bozuyuk2022,Ham2021}, when the channel cross section becomes sufficiently small. Interestingly, although at low \Rey\ the rotation frequency is expected to only set the time scale of motion, experiments suggest that the behavior of microrollers under confinement may depend on the rotation frequency. For example, under the same degree of confinement, microrollers rotating at higher frequencies experience a greater reduction in translational velocity~\citep{Bozuyuk2022}. This apparent discrepancy likely arises from the frequency dependence of the particle-surface gap distance~\citep{Disharoon2019}.  

To understand how microroller motility varies with confinement, previous studies typically compute the hydrodynamic forces acting on the microroller by directly solving the Stokes equations, while assuming a fixed microroller position~\citep{Fang2020,Bozuyuk2021,Bozuyuk2022}. Such approaches require detailed analyses of the pressure and flow fields specific to each geometry. Importantly, the observation of velocity reversal across different geometries suggests the existence of a universal underlying mechanism; however, a unified understanding remains elusive. In addition, although modeling a microroller as a rotlet singularity has been shown to successfully capture several collective behaviors~\citep[see e.g.,][]{Driscoll2017}, it remains unclear how the rotlet representation describes the dynamics of an individual microroller under confinement. Understanding this mechanism may provide insights for designing microfluidic systems that control both the dynamics of individual microrollers and their collective motion.

In this work, we develop a systematic scaling framework to investigate the dynamics of a microroller in confined geometries. Our analysis is complemented by experiments and numerical simulations. Using the force-free condition for a freely translating microroller, we derive the functional dependence of the translational velocity on parameters characterizing the degree of confinement and extend the analysis to a variety of channel geometries. The resulting scaling predictions show excellent agreement with both experimental measurements and simulation results, demonstrating that the rotlet flow is a key mechanism underlying velocity reduction and reversal in confined channels.

The paper is organized as follows. In \S~\ref{sec:Experiments}, we present experimental results on velocity reduction and reversal in constricted and rectangular channels. In \S~\ref{sec:Numerical}, we describe our simulation model based on a boundary-element method and validate it by comparing the numerical results with the asymptotic solutions derived in Appendix~\ref{sec:AppendixA}. In \S~\ref{sec:Scaling}, we develop the scaling framework, beginning with the case of a single bounding surface and extending it to constricted, rectangular, and cylindrical channels. 

\section{Experiments}\label{sec:Experiments}
\begin{figure}
\centerline{\includegraphics[width=1.0\textwidth]{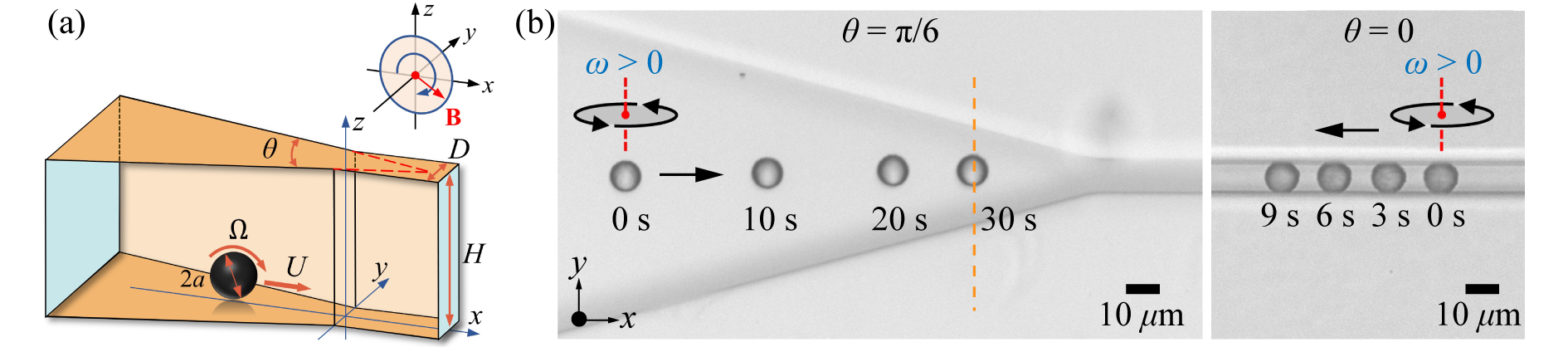}}
\caption{(a) Schematic illustration of a magnetic particle moving in a constricted microchannel driven by a rotating magnetic field $\boldsymbol{B}$. (b) Sequential snapshots at equal time intervals showing the translation of a magnetic particle in a constricted channel ($\theta = \pi/6$, $D = 10.7$ $\mu$m; left) and a straight channel ($D = 10.3$ $\mu$m; right). The orange dashed line marks the stopping position of the particle in the constricted channel.}
\label{fig:1}
\end{figure}
The microfluidic channels shown in Fig.~\ref{fig:1} are designed and fabricated by standard photolithography protocols. Two types of channels are considered in the experiments: constricted channels with gradually decreasing cross-sections and straight rectangular channels. The geometry of the constricted microchannel [Fig.~\ref{fig:1}(a)] is characterized by the opening angle $\theta$ and the width $D$ of the downstream straight channel. The rectangular channel is characterized by the width $D$. The depth $H$ of both channels is $H = 30\ \mu\mathrm{m}$. By comparison with measurements in a microwell with a free top surface, we find that the top plate of the microchannel has only a minor effect on the microroller velocity.

The magnetic field is generated using a customized triaxial coil system. We program the generator to produce a magnetic field rotating around the $y$ axis, $\boldsymbol{B}(t) = B_0 (\cos\omega t, 0, \sin\omega t)$, where $B_0$ and $\omega$ are the field strength and angular frequency, respectively. In our experiments, the typical field strength is $B_0 \approx 4\ \mathrm{mT}$, and $\omega$ is below the critical step-out frequency, beyond which the particle no longer rotates synchronously with $\boldsymbol{B}$. 

We use a diluted aqueous suspension of ferromagnetic microspheres, composed of a polystyrene matrix doped with nanometric iron oxide grains (Nantong Zhichuan Microsphere Biotechnology Co., Ltd.). The particle radius is $a$ = 5.0 $\pm$ 0.7 $\mu$m. The balance between buoyancy and the electrostatic repulsion from the charged boundary keeps the particles from touching the bottom plate. The gap distance between the particle surface and the plate, $h$, is much smaller than $a$ and is typically on the order of tens of nanometers~\citep{Disharoon2019}. We initially control the particle positions to be close to the channel centerline. Driven by the rotating magnetic field, particles rotate with angular velocity $\boldsymbol{\Omega} = \Omega \hat{\boldsymbol{y}}$, where $\Omega = \omega$ in our experiments, and simultaneously translate with velocity $\boldsymbol{U} = U\hat{\boldsymbol{x}}$. We record their motions using a CMOS camera (Teledyne Photometrics, Prime BSI) and extract the trajectories from recorded image sequences using a customized algorithm. 

\subsection{Experimental results}
\begin{figure}
\centerline{\includegraphics[width=1.0\textwidth]{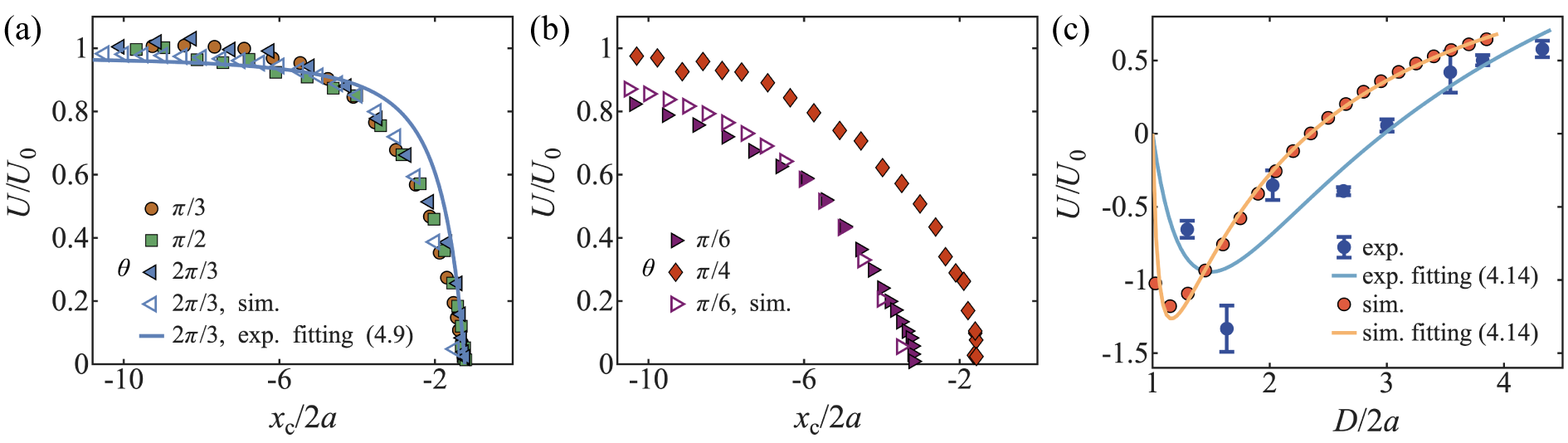}}
\caption{(a) Translational velocities $U/U_0$ in constricted channels as a function of the center-of-mass position of the microroller $x_\mathrm{c}/2a$. The opening angles are $\theta$ = $\pi/3$, $\pi/2$ and $2\pi/3$, and the width of the downstream straight channel is $D \approx 10.7\ \mu\mathrm{m}$. The solid line is the best fit of the scaling relation (\ref{U_scaling_2}) to the experimental data for $\theta = 2\pi/3$. (b) $U/U_0$ as a function of $x_\mathrm{c}/2a$ for smaller opening angles $\theta$ = $\pi/6$ and $\pi/4$. (c) $U/U_0$ as a function of $D/2a$ for microrollers moving in straight channels. Error bars represent uncertainties caused by variations in particle size and relative position to the side walls. The solid lines represent the best fits of the scaling relation (\ref{U_scaling_3}). The angular velocity of the microrollers in experiments is $\Omega = 8\pi\ \mathrm{rad}/s$.}
\label{fig:2}
\end{figure}

The translational velocity of a microroller is much smaller than the pure rolling velocity $\Omega a$. For instance, when the microroller is far from the side walls and only the bottom plate is in effect, the translational velocity $U_0 = 6.3\ \mu$m/s for $\Omega = 8\pi$ rad/s and $a = 4.6$ $\mu$m. Different microrollers exhibit slightly different translational velocities.

We first examine the microroller motion in the constricted channel. A typical example is shown in Fig.~\ref{fig:1}(b, left panel) for $\theta = \pi/6$ and $D = 10.7$ $\mu\mathrm{m}$. The microroller gradually slows down and eventually stops near the entrance of the downstream straight channel (see Movie 1 in the Supplemental Material). To quantify this behavior, we plot in Figs.~\ref{fig:2}(a) and \ref{fig:2}(b) the translational velocity $U$, normalized by $U_0$, as a function of the normalized center-of-mass position along the $x$ axis, $x_\mathrm{c}/2a$. The origin of the coordinate system is located at the entrance of the straight channel. For different values of $\theta$, $U$ exhibits a consistent decrease. The velocity of the microroller decreases more rapidly as it approaches the straight channel. Notably, for large opening angles $\theta \gtrsim \pi/3$, the data collapse onto a single trend, and microrollers stop approximately at the same position around $x \approx -3a$ [Fig.~\ref{fig:2}(a)]. This suggests that, for large $\theta$, the two side walls of the constricted region have a negligible effect, and the observed slowing down is mainly caused by the straight channel. For smaller opening angles, e.g., $\theta = \pi/4$ and $\pi/6$ [Fig.~\ref{fig:2}(b)], the side walls become closer and begin to influence the microroller dynamics. As a result, $U$ demonstrates an earlier decrease, with the stopping position shifting away from the entrance. By varying the width of the straight channel $D$, we find that a stopping position exists for $D \lesssim 30$ $\mu$m, whereas increasing $D$ beyond this threshold allows microrollers to enter the straight channel.

We then investigate the dynamics of microrollers moving in the straight channel. As shown in Fig.~\ref{fig:1}(b) right panel, in channels with $D$ close to the particle diameter, e.g., $D = 10.3$ $\mu$m, the microroller translates backward, opposite to the rolling direction (Movie 2). In channels with different widths, we measure the average translational velocity over multiple trials with different microrollers. As $D$ increases, the backward velocity first increases in magnitude, reaching $U/U_0 \approx -1.4$ at $D/2a \approx 1.6$ [Fig.~\ref{fig:2}(c)], and then decreases to zero around $D/2a \approx 3.0$. Further increasing $D$ recovers the normal forward motion.

\section{Numerical simulation}\label{sec:Numerical}
\subsection{Simulation model}
\begin{figure}
\centerline{\includegraphics[width=1.0\textwidth]{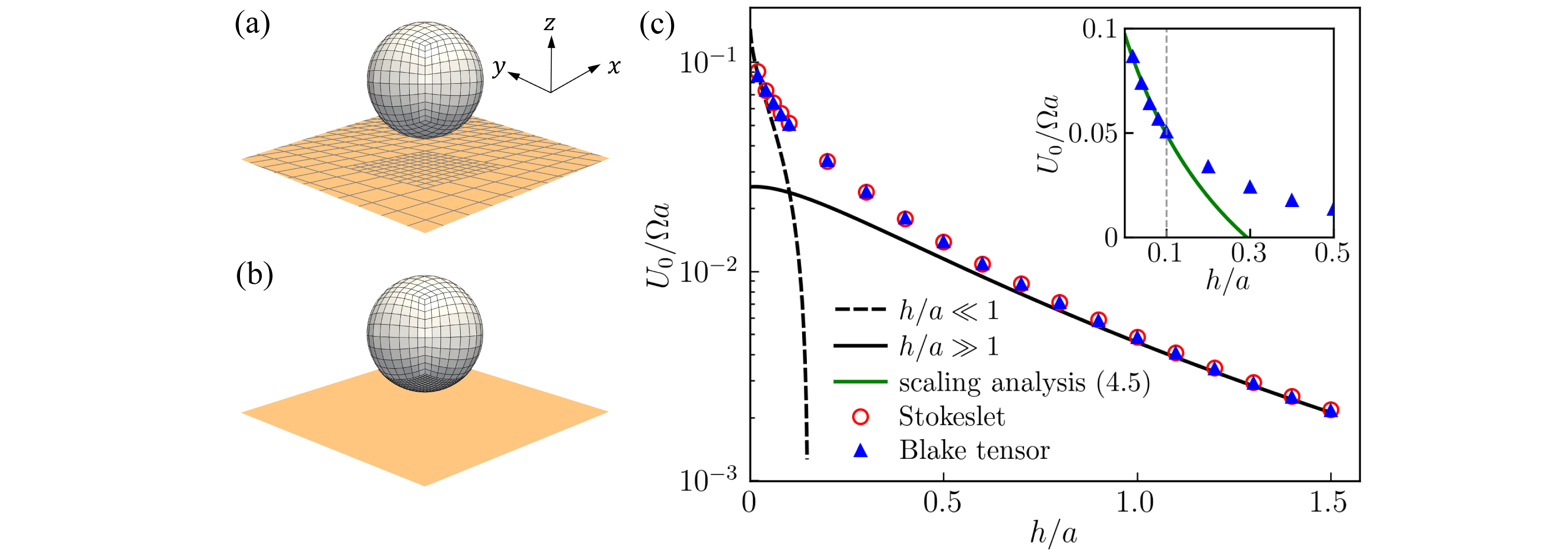}}
\caption{(a),(b) Computational meshes for a microroller moving near a no-slip boundary using (a) the free-space Stokeslet and (b) the Blake tensor. (c) Comparison of the numerical results for $U_0$ obtained from the boundary element method with the asymptotic solutions in the limits of $h/a \ll 1$ (\ref{U_asympt1}) and $h/a \gg 1$ (\ref{U_asympt2}). The inset shows the best fit of the scaling relation for $U_0 (h)$ (\ref{U_approx_1}) to the numerical results for $h/a < 0.1$.}
\label{fig:3}
\end{figure}
We simulate microroller dynamics in confined microchannels using a hybrid boundary element and regularized Stokeslet method~\citep{Smith2009,Cortez2005}. Using a specific cutoff function, the regularized Stokeslet is expressed as
\begin{equation}\label{reg_stokeselt}
\mathsfbi{G}_{\epsilon} (\boldsymbol{x}_0, \boldsymbol{x}) = \frac{\mathsfbi{I}(R^2 + 2 \epsilon^2) + \boldsymbol{R}\boldsymbol{R}}{R_{\epsilon}^3},
\end{equation}
where $\epsilon$ is the regularization parameter, $\boldsymbol{R} = \boldsymbol{x}_0-\boldsymbol{x}$, and $R_{\epsilon} = \sqrt{R^2 + \epsilon^2}$. The motion of a microroller under confinement is represented by distributions of regularized Stokeslets on its surface $S_\mathrm{p}$ and the channel walls $S_\mathrm{w}$. The disturbance velocity at a point $\boldsymbol{x}_0$ in the fluid domain is given by
\begin{equation}\label{disturbance_flow}
\boldsymbol{u}(\boldsymbol{x}_{0}) = \frac{1}{8 \pi \mu}\int_{S_\mathrm{p} + S_\mathrm{w}}\mathsfbi{G}_{\epsilon}(\boldsymbol{x}_{0}, \boldsymbol{x}) \cdot \boldsymbol{f}(\boldsymbol{x})\, \mathrm{d}\boldsymbol{x}.
\end{equation}
On the boundaries, the disturbance velocity satisfies the following no-slip conditions,
\begin{gather}
\boldsymbol{u}(\boldsymbol{x}_{0}) = U\hat{\boldsymbol{x}} + \boldsymbol{\Omega} \times(\boldsymbol{x}_{0} - \boldsymbol{r}_\mathrm{c}),\quad \text{for }\boldsymbol{x}_0 \in S_\mathrm{p}, \label{boundary1} \\
\boldsymbol{u}(\boldsymbol{x}_{0})=0,\quad \text{for }\boldsymbol{x}_{0}\in S_\mathrm{w},
\end{gather}
where the center of the microroller is $\boldsymbol{r}_\mathrm{c} = (x_\mathrm{c}, 0, z_\mathrm{c})$ and its angular velocity is $\boldsymbol{\Omega} = \Omega \hat{\boldsymbol{y}}$. Finally, since only an external torque is applied to the microroller, the net hydrodynamic force is zero,
\begin{equation}\label{force_free}
\int_{S_\mathrm{p}}\boldsymbol{f}(\boldsymbol{x})\, \mathrm{d} \boldsymbol{x} = 0.
\end{equation}

We nondimensionalize the equations by scaling length by $a$, time by $T = 2\pi/\Omega$, and force by $f_0 = \mu a^2 /T$. The integral formulations (\ref{disturbance_flow})--(\ref{force_free}) are discretized into a system of linear equations for the unknowns $\boldsymbol{f}$ and $U$. Equation~(\ref{disturbance_flow}) is approximated as
\begin{equation}\label{discretization}
\boldsymbol{u} (\boldsymbol{x}_0) = \frac{1}{8\pi\mu} \sum_{n=1}^N \boldsymbol{f}^n\cdot \int_{\delta S^n} \mathsfbi{G}_{\epsilon}(\boldsymbol{x}_0, \boldsymbol{x}) \,\mathrm{d} \boldsymbol{x},
\end{equation}
where $N$ is the number of surface elements $\delta S^n$, each associated with a force density $\boldsymbol{f}^n$ assumed to act at the center of the element. In (\ref{discretization}), the integral of $\mathsfbi{G}_{\epsilon}$ over each element $\delta S^n$ is evaluated using Gauss–Legendre quadrature with $m\times m$ points. 

We validate our numerical method by performing simulations of a microroller moving above an infinite plate. The surfaces of the microroller and the plate are discretized as illustrated in Fig.~\ref{fig:3}(a). A finer mesh with an average element size of $\delta s = 0.125 a$ is used on the plate around the closest contact point and over the entire sphere surface. On the rest of the plate, $\delta s = 0.4 a$. The number of quadrature points is $m = 12$, and the regularization parameter is $\epsilon = \delta s/m$. Figure~\ref{fig:3}(c) shows that the translational velocity $U_0$ increases monotonically as the minimum distance $h$ decreases. We also perform simulations using the Blake tensor~\citep{Blake1971}, which automatically satisfies the no-slip condition at the plate. Therefore, only the microroller surface needs to be discretized [Fig.~\ref{fig:3}(b)]. For $h < 0.1 a$, a finer mesh with an average size of $\delta s = 0.02 a$ is used around the nearest contact point. The results match closely with those obtained using the free-space Stokeslet [Fig.~\ref{fig:3}(c)]. In Appendix~\ref{sec:AppendixA}, we derive asymptotic expressions for $U_0$ in the limits $h/a \ll 1$ and $h/a \gg 1$. Excellent agreement with simulation results is observed in both limits [Fig.~\ref{fig:3}(c)], validating our simulation model. 

\subsection{Simulation results}
We first perform simulations in the constricted channels. The top plate is neglected and the channel depth is set sufficiently large, $H=10 a$. The gap distance $h$ is fixed at $h \approx 0.12a$ in our simulations. This value is chosen such that the computed translational velocity approximately matches the experimentally measured velocity $U_0$. For both small and large opening angles, the computed velocities at different positions are in excellent agreement with the experimental results [Figs.~\ref{fig:2}(a) and \ref{fig:2}(b)], suggesting that the slowing of microrollers in constricted channels originates from hydrodynamic interactions with boundaries.

In straight channels, as $D$ decreases, simulations reproduce the velocity reversal at a critical channel width [Fig.~\ref{fig:2}(c)]. The magnitude of the backward velocity also reaches a maximum before decreasing as $D$ is further reduced. Although the overall trend of $U(D)$ is consistent with the experimental results, a discrepancy in $U$ is observed for a given $D$. This may result from variations in particle sizes and position in the experiments.

\begin{figure}
\centerline{\includegraphics[width=1.0\textwidth]{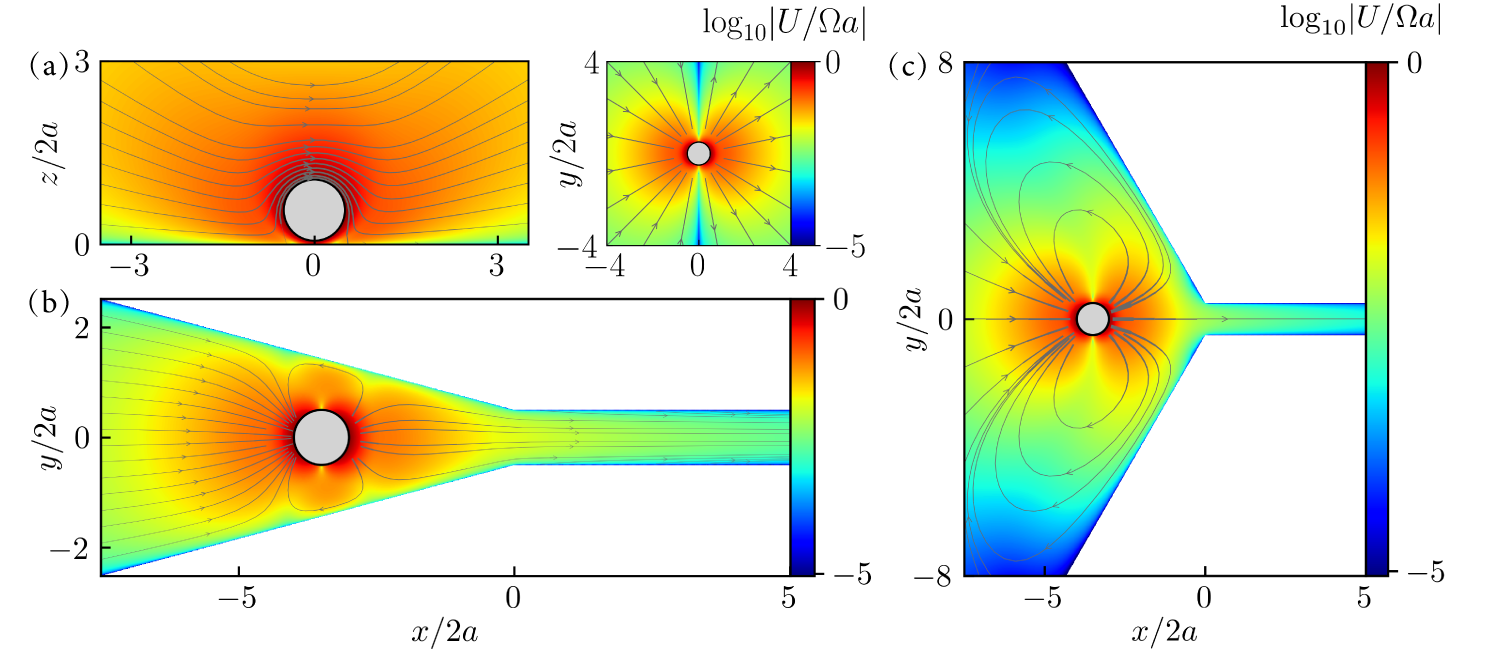}}
\caption{Disturbance flow fields around a microroller. (a) Flow fields in the $x$-$z$ and $x$-$y$ planes for a microroller near an infinite plate. (b), (c) Flow fields in the $x$-$y$ plane for constricted channels with opening angles (b) $\theta = \pi/6$ and (c) $\theta = 2\pi/3$. All cross-sectional planes pass through the microroller center.}
\label{fig:4}
\end{figure}

To gain physical intuition, we plot the instantaneous flow fields around the microroller in different geometries obtained from simulations (Fig.~\ref{fig:4}). The rotational motion not only disturbs the nearby fluid but also entrains a large-scale directional flow along the $x$ axis [Fig.~\ref{fig:4}(a)]. In constricted channels with both small ($\theta = \pi/6$) and large ($\theta = 2\pi/3$) opening angles, such directional flows extend deep into the downstream straight channel [Figs.~\ref{fig:4}(b) and \ref{fig:4}(c)]. Therefore, the slowing down of microrollers likely arises from the additional flow resistance generated in the straight channels. Meanwhile, for $\theta = \pi/6$, the side walls strongly distort the flow streamlines, leading to quantitatively different velocity trends compared with those at large $\theta$ [Figs.~\ref{fig:2}(a) and \ref{fig:2}(b)]. To reveal the mechanisms underlying velocity reduction and reversal, we next perform the scaling analysis based on the force-free condition (\ref{force_free}).

\section{Scaling analysis and mechanisms}\label{sec:Scaling}
\subsection{Semi-infinite space}\label{sec:semi-infinite}
We first consider the case in which only the bottom plate is present. The angular velocity of the microroller is $\boldsymbol{\Omega} = \Omega \hat{\boldsymbol{y}}$. Since an external magnetic torque is applied on the microroller, we approximate the far-field disturbance flow as that generated by a rotlet singularity. In the presence of a no-slip boundary, the flow velocity at a spatial position $\boldsymbol{x}_0$ generated by a rotlet positioned at $\boldsymbol{r}_\mathrm{c} = (x_\mathrm{c},0,z_\mathrm{c})$ is given by~\citep{Blake1974}
\begin{equation}\label{rotlet_flow}
\begin{aligned}
8\pi\mu u_i(\boldsymbol{x}_0, \boldsymbol{r}_\mathrm{c}) &= \frac{\epsilon_{ijk}\tau_j  R_k}{R^3} + \left[-\frac{\epsilon_{ijk} \tau_j \tilde{R}_k}{\tilde{R}^3} + 2 z_\mathrm{c} \epsilon_{kjz} \tau_j \left(\frac{\delta_{ik}}{\tilde{R}^3} - \frac{3\tilde{R}_i \tilde{R}_k}{\tilde{R}^5}\right) \right. \\
&\quad \left. +\, 6 \epsilon_{kjz} \frac{\tau_j \tilde{R}_i \tilde{R}_k \tilde{R}_z}{\tilde{R}^5}\right], \quad \text{for }i,j,k = x,y,z.
\end{aligned}
\end{equation}
where $\boldsymbol{R} = \boldsymbol{x}_0 - \boldsymbol{r}_\mathrm{c}$, $\tilde{\boldsymbol{R}} = \boldsymbol{x}_0 - (\boldsymbol{r}_\mathrm{c} - 2 z_\mathrm{c} \hat{\boldsymbol{z}})$, and $\boldsymbol{\tau} = 8\pi\mu a^3 \boldsymbol{\Omega}$. Setting $z_\mathrm{c} = h+a$ and the evaluation position $\boldsymbol{x}_0 = (x_0, 0, h+a)$, we obtain the flow velocity along the horizontal line passing through the center of the microroller. In the limit $|x_0 - x_\mathrm{c}| \gg z_\mathrm{c}$, the $x$-component velocity is given by
\begin{equation}\label{far_velocity}
u_x (x_0, x_\mathrm{c}) \approx 6\Omega a^{3} (h+a) \frac{1}{|x_0 - x_\mathrm{c}|^{3}}.
\end{equation}

The translational motion of the microroller arises from an asymmetry in the viscous stress acting on the upper and lower regions of its surface. If the microroller is constrained to rotate in place, its lower surface experiences a viscous stress from the thin fluid layer between the microroller and the bottom plate, which scales as $\sigma_{\mathrm{L}} \sim \eta \Omega a/h$. The shear between the upper surface and the surrounding fluid generates a viscous stress that scales as $\sigma_{\mathrm{U},\text{near}} \sim \eta \Omega a/a = \eta \Omega$, where the disturbance flow due to the rotational motion is considered to vary over a characteristic scale on the order of $a$ in the $z$ direction. Meanwhile, the motion of the upper surface entrains a far-field flow along the $+x$ direction [Fig.~\ref{fig:4}(a)], which is approximated by (\ref{rotlet_flow}). This far-field flow generates a viscous shear stress on the bottom plate. Because the plate is held fixed by an external support, this wall-induced stress effectively contributes to the hydrodynamic resistance experienced by the upper surface of the microroller. This additional contribution is estimated using (\ref{far_velocity}) in the limit $h \ll a$ as
\begin{equation}\label{sigma_U_far}
\begin{aligned}
\sigma_{\mathrm{U},\text{far}} & \sim \frac{1}{a^2} \left(\int^{x_\mathrm{c} - a}_{-\infty} + \int_{x_\mathrm{c} + a}^{\infty} \right) \eta\frac{u_x(x_0, x_\mathrm{c})}{a} a\, \mathrm{d}x_0 \\
& \sim  \eta \Omega,
\end{aligned}
\end{equation}
where $a\mathrm{d}x_0$ represents a differential area element on the bottom plate, and the near-field region adjacent to the microroller is excluded from the integration. Because $h \ll a$, $\sigma_{\mathrm{L}}$ is not balanced by the sum of $\sigma_{\mathrm{U,near}}$ and $\sigma_{\mathrm{U,far}}$, leading to a net propulsion force along the $+x$ direction. 

If the microroller is free to translate, the force-free constraint (\ref{force_free}) can be satisfied with a nonzero translational velocity $U_0\hat{\mathbf{x}}$. The stress $\sigma_{\mathrm{L}}$ is then reduced to $\sigma_{\mathrm{L}} \sim \eta (\Omega a - \kappa U_0)/h$, where $\kappa$ is a geometrical constant characterizing the relative contributions of rotational and translational velocities to the shear. In contrast, $\sigma_{\mathrm{U,near}}$ is increased to $\sigma_{\mathrm{U,near}} \sim \eta (\Omega a + \kappa U_0)/a$. Balancing $\sigma_{\mathrm{L}}$ with $\sigma_{\mathrm{U,near}}$ and $\sigma_{\mathrm{U,far}}$ yields
\begin{equation}\label{stress_balance1}
\eta \frac{\Omega a - \kappa U_0}{h} \approx \alpha_1\eta \frac{\Omega a + \kappa U_0}{a} + \alpha_2\eta \Omega.
\end{equation}
Here, we introduce two dimensionless prefactors $\alpha_1$ and $\alpha_2$ to represent the relative magnitudes of the corresponding terms. Equation~(\ref{stress_balance1}) forms the basis for the scaling analysis in more complex geometries. Solving (\ref{stress_balance1}) for $U_0$, we obtain
\begin{equation}\label{U_approx_1}
U_0(h) \sim \Omega a \frac{ah^{-1} - \alpha_1 - \alpha_2}{ah^{-1} + \alpha_1},
\end{equation}
where $\kappa$ has been absorbed into the overall prefactor of the scaling relation. Equation (\ref{U_approx_1}) is valid in the limit $h \ll a$. As shown in Fig.~\ref{fig:3}(c) inset, (\ref{U_approx_1}) closely fits the numerical results for small $h$. The fitted parameters are $\alpha_1 = 3.0$ and $\alpha_2 = 0.5$, indicating that $\sigma_{\mathrm{U,near}}$ is the dominant contribution to the viscous stress on the upper surface, while the rotlet-induced stress $\sigma_{\mathrm{U,far}}$ remains non-negligible. 

\subsection{Constricted channel}\label{sec:constricted_channel}
For the constricted channels considered in our experiments, the far-field flow induced by microroller rotation generates viscous stresses on the side walls of both the constricted region and the downstream straight channels. These stresses effectively increases the stress acting on the upper surface of the microroller and thereby reducing the translational velocity. We focus on the case of large opening angles $\theta$, for which the effect of the side walls in the constricted region is minor [Fig.~\ref{fig:2}(a)]. The viscous stress arising from the shear between the rotlet flow and the side walls of the straight channel scales as 
\begin{equation}\label{sigma_U_side}
\frac{1}{a^2} \int^{\infty}_{0} \eta \frac{u_x (x_0, x_\mathrm{c})}{D} a\, \mathrm{d}x_0 \sim \eta \Omega a^{3} \frac{1}{Dx_\mathrm{c}^{2}},
\end{equation}
where $a\mathrm{d}x_0$ represents a differential surface element on the side walls, and the integration extends from the entrance of the straight channel to infinity. Including this additional contribution in the stress balance, (\ref{stress_balance1}) becomes
\begin{equation}\label{stress_balance2}
\eta\frac{\Omega a - \kappa U}{h} \approx \alpha_1\eta\frac{\Omega a + \kappa U}{a} + \alpha_2 \eta \Omega + \alpha_3\eta\Omega a^{3} \frac{1}{D x_\mathrm{c}^{2}},
\end{equation}
where the dimensionless prefactors $\alpha_1$, $\alpha_2$, and $\alpha_3$ weight the respective stress contributions. Solving (\ref{stress_balance2}) for $U$ yields
\begin{equation}\label{U_approx_2}
U \sim \Omega a \frac{ah^{-1} - \alpha_1 - \alpha_2 - \alpha_3 a^3/(D x_\mathrm{c}^2) }{ah^{-1} + \alpha_1}.
\end{equation}
Because our primary interest is in the dependence of $U$ on $x_\mathrm{c}$, and $h$ can be treated as a constant, (\ref{U_approx_2}) then reduces to the following scaling relation,
\begin{equation}\label{U_scaling_2}
U(x_\mathrm{c}) \sim \Omega a\left(1 - \beta \frac{a^3}{D x_\mathrm{c}^{2}}\right),
\end{equation}
where $\beta$ is a dimensionless parameter determined by fitting the experimental or simulation data. 

As shown in Fig.~\ref{fig:2}(a), (\ref{U_scaling_2}) agrees well with both the simulation and experimental data. The best-fit parameter is $\beta = 18.0$. This agreement validates (\ref{far_velocity}) as an approximation for the far-field disturbance flow. The viscous stress generated by the rotlet flow along the side walls of the straight channel is therefore the primary mechanism underlying the observed velocity reduction.

\subsection{Rectangular channel}\label{sec:rectangular_channel}
For a microroller moving in a straight rectangular channel with a large depth, as in the experiments, we neglect the effect of the top plate. The far-field rotlet flow generates viscous stresses on the sidewalls, which contribute an additional resistance to microroller translation. Using (\ref{far_velocity}), this contribution is estimated as
\begin{equation}\label{stress_side_walls}
\frac{1}{a^2} \left(\int^{x_\mathrm{c} - a}_{-\infty} + \int_{x_\mathrm{c} + a}^{\infty} \right) \eta\frac{u_x(x_0, x_\mathrm{c})}{D} a\, \mathrm{d}x_0 \sim \eta \Omega \frac{a}{D},
\end{equation}
where $a\mathrm{d}x_0$ represents a differential area element on the side walls. Equation~(\ref{stress_side_walls}) is similar to (\ref{sigma_U_far}), except that the characteristic scale of velocity variation is replaced by the channel width $D$. In addition, when the microroller translates, the shear in the narrow gaps between its lateral surfaces and the two side walls generates a viscous stress that scales as 
\begin{equation}\label{lateral_shear1}
\eta U/(D - 2a).
\end{equation}
Since $U \ll \Omega a$, this term is negligible for $D \gg 2a$, but becomes dominant when $D$ approaches $2a$. Including both contributions, the stress balance becomes 
\begin{equation}\label{stress_balance3}
\eta \frac{\Omega a - \kappa U}{h} \approx \alpha_1 \eta \frac{\Omega a + \kappa U}{a} + \alpha_2 \eta \Omega + \alpha_3 \eta \Omega \frac{a}{D} + \alpha_4 \eta \frac{U}{D - 2a}.
\end{equation}
where the dimensionless prefactors $\alpha_1$, $\alpha_2$, $\alpha_3$, and $\alpha_4$ weight the respective stress contributions. Solving (\ref{stress_balance3}) for $U$ yields
\begin{equation}\label{U_approx_3}
U \sim \Omega a \frac{a h^{-1} - \alpha_1 - \alpha_2 - \alpha_3 aD^{-1}}{a h^{-1} + \alpha_1 + \kappa^{-1}\alpha_4 a (D-2a)^{-1}}.
\end{equation}
The dependence of $U$ on $D$ can be recast as
\begin{equation}\label{U_scaling_3}
U(D) \sim \Omega a \frac{1 - \beta_1 a D^{-1}}{1 + \beta_2 a (D-2a)^{-1}},
\end{equation}
where $\beta_1$ and $\beta_2$ are two dimensionless fitting parameters.

Figure~\ref{fig:2}(c) compares the scaling prediction with both the simulation and experimental results. The fitted parameters are $\beta_1 = 4.7$ and $\beta_2 = 0.08$. Despite the simplifications, (\ref{U_scaling_3}) successfully captures the functional dependence of $U$ on $D$. The additional viscous stress generated by the far-field rotlet flow (\ref{stress_side_walls}) is the primary mechanism responsible for the observed velocity reversal. This can be seen from the numerator of (\ref{U_scaling_3}), which changes sign when $D$ falls below the critical value $\beta_1 a$. Additionally, (\ref{U_scaling_3}) also guarantees that $U(D)$ vanishes as $D$ approaches $2a$, because the term $\beta_2 a (D-2a)^{-1}$ in the denominator diverges. In this limit, the translational resistance generated by the shear between the lateral particle surfaces and the sidewalls, (\ref{lateral_shear1}), dominates and suppresses microroller translation.

\subsection{Cylindrical channel}
\begin{figure}
\centerline{\includegraphics[width=1.0\textwidth]{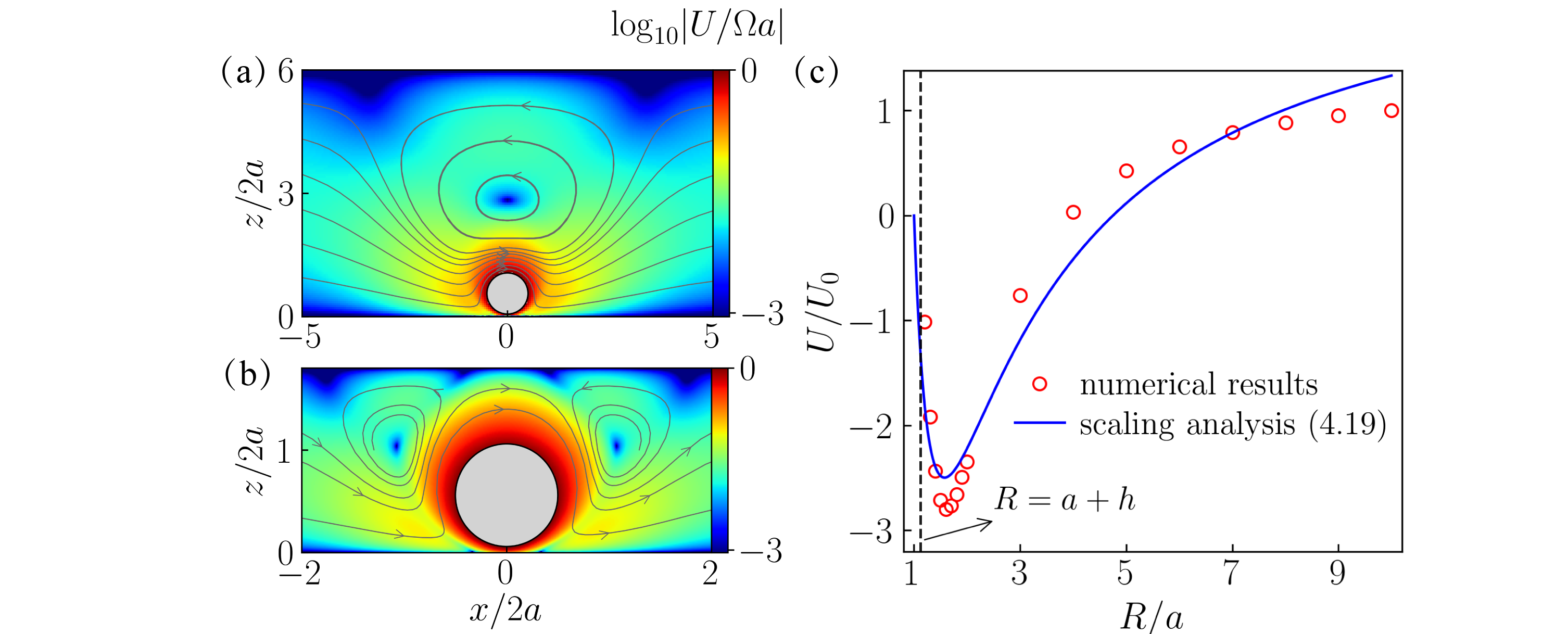}}
\caption{(a), (b) Disturbance flow fields in the $x$-$z$ plane for a microroller moving in a cylindrical channel of radius (a) $R = 6.0 a$ and (b) $R = 1.8 a$. (c) $U/U_0$ as a function of the channel radius $R/a$. The solid curve shows the best fit of the scaling relation (\ref{U_scaling_4}). The dashed line indicates the radius $R = a+h$, at which the translational velocity is expected to vanish by symmetry. In the numerical simulations, the cylinder length is set to $40 a$, which is sufficiently long to approximate an infinitely long channel.}
\label{fig:5}
\end{figure}

To gain physical intuition, we first compute the disturbance flow field using the boundary element method (\S~\ref{sec:Numerical}). An example is shown in Fig.~\ref{fig:5}(a) for a cylindrical channel of radius $R = 6a$. A notable feature of the flow field is the existence of a large vortex between the microroller and the upper channel wall. A strong shear flow extends across the region between the upper surface of the microroller and the vortex center. As $R$ decreases to $R = 1.8a$, the large-scale unidirectional flow in the $x$ direction persists. Meanwhile, the single vortex splits into two vortices located in front of and behind the microroller. In both cases, the characteristic length scale of the shear flow near the upper surface of the microroller remains on the order of $a$. Therefore, the corresponding viscous stress scales as $\eta (\Omega a + \kappa U)/a$. However, this estimate is expected to break down under extremely tight confinement, $R-a \sim h$, where the characteristic length scale becomes $R-a$. 

The far-field rotlet flow generates an additional viscous stress on the cylindrical wall, which contributes to the stress balance as
\begin{equation}\label{cylinder_rotlet}
\frac{1}{a^2} \left(\int^{x_\mathrm{c} - a}_{-\infty} + \int_{x_\mathrm{c} + a}^{\infty} \right) \eta\frac{u_x(x_0, x_\mathrm{c})}{R} a\, \mathrm{d}x_0 \sim \eta \Omega \frac{a}{R}.
\end{equation}
Unlike the rectangular channel, we do not distinguish between the bottom wall and the side walls in the cylindrical geometry. Finally, the viscous stress associated with microroller translation scales as
\begin{equation}\label{lateral_shear2}
\eta U/(R-a), 
\end{equation}
which becomes important only $R$ is close to $a$. Collecting these contributions, the stress balance for a cylindrical channel can be written as
\begin{equation}\label{stress_balance4}
\eta \frac{\Omega a - \kappa U}{h} \approx \alpha_1 \eta \frac{\Omega a + \kappa U}{a} + \alpha_2 \eta \Omega \frac{a}{R} + \alpha_3 \eta \frac{U}{R-a},
\end{equation}
where the dimensionless prefactors $\alpha_1$, $\alpha_2$, and $\alpha_3$ weight the respective stress contributions. Solving (\ref{stress_balance4}) for $U$ yields
\begin{equation}\label{U_approx_4}
U \sim \Omega a \frac{ah^{-1} - \alpha_1 - \alpha_2 a R^{-1}}{ah^{-1} + \alpha_1 + \kappa^{-1}\alpha_3 a (R-a)^{-1}},
\end{equation}
Focusing on the dependence of $U$ on $R$, (\ref{U_approx_4}) can be further simplified to
\begin{equation}\label{U_scaling_4}
U(R) \sim \Omega a \frac{1 - \beta_1 a R^{-1}}{1 + \beta_2 a (R - a)^{-1}},
\end{equation}
which is valid for $a \gg h$ and $R-a \gg h$. Despite the difference in geometry, (\ref{U_scaling_4}) has a similar structure to (\ref{U_scaling_3}).

Figure~\ref{fig:5}(c) shows that (\ref{U_scaling_4}) agrees well with the simulation results. The fitted parameters are $\beta_1 = 4.8$ and $\beta_2 = 0.7$. As $R$ decreases, the term $\beta_1 a R^{-1}$ in the numerator of (\ref{U_scaling_4}) eventually becomes larger than 1. Therefore, as in the rectangular channel, velocity reversal in the cylindrical channel is also caused by the additional viscous stress generated by the far-field rotlet flow (\ref{cylinder_rotlet}). In addition, as $R$ approaches $a$, the translational resistance (\ref{lateral_shear2}) diverges, causing the magnitude of the backward velocity to decrease rapidly toward zero. However, because the derivation of (\ref{stress_balance4}) assumes $h \ll a$, the resulting scaling relation predicts that $U(R)$ vanishes at $R = a$ rather than at the physical limit $R = a+h$. Nevertheless, Fig.~\ref{fig:5}(c) shows that (\ref{U_scaling_4}) provides a reasonable approximation to the simulation results for $R > a$.

\section{Discussion and Conclusion}
In this work, we investigated the dynamics of a microroller moving in confined microchannels by combining experiments, numerical simulations, and scaling analysis. To develop a scaling framework, we decompose the disturbance flow generated by a microroller into far-field and near-field components, representing the rotlet flow and the shear flow associated with the relative motion between the microroller and the surrounding fluid, respectively. This decomposition enables the analysis to be generalized to different geometries and to highlight the distinct roles of the two flow components. The scaling predictions agree well with both simulations and experiments. We demonstrate that the viscous stress generated by the rotlet flow governs both velocity reduction and reversal across different geometries, while the translational resistance associated with the near-field shear flow suppresses the microroller velocity under tight confinement.

Our results may provide guidance for controlling microroller dynamics in confined fluid environments. For example, by designing the ratio of the microroller size to the channel width, one may enable microrollers to access confined spaces or to translate with a desired velocity and direction. In addition, our results suggest that spheroidal particles are preferable to spherical particles for locomotion in confined spaces. This is because, when compared with a sphere of radius equal to the spheroid's semi-major axis, a spheroid has much smaller rotational resistance coefficients about both the major and minor axes~\citep{Kim13}. As a result, the spheroid generates a weaker rotlet flow and therefore a smaller far-field viscous stress. Experiments have indeed confirmed that elongated particles outperform spherical ones in confined channels~\citep{Bozuyuk2022}.

Interestingly, in experiments with constricted channels, we find that microrollers driven at a low angular velocity $\Omega = \pi$ rad/s are able to slowly creep into the straight channel, whereas those with higher angular velocities cannot. This may arise from variations in the gap distance between the microroller and the bottom wall at different rotational frequencies~\citep{Disharoon2019}, but the exact mechanism remains unclear. Several mechanisms can be exploited to modify the particle-wall gap distance and thereby control microroller motility, including manipulation of the electric double layer and motion near deformable boundaries~\citep{Rallabandi24,Hu2023_2} or boundaries with nonuniform slippage~\citep{Rinehart2020}, where nontrivial lift forces can arise. In this work, we have focused on Newtonian fluids. In viscoelastic fluids, velocity reversal can occur without confinement~\citep{He2026}. A natural extension of this work is to investigate how confinement and fluid viscoelasticity together affect microroller motility.

\backsection[Funding]{S. Hu acknowledges support from the National Natural Science Foundation of China (NSFC grant no. 12504234) and the Fundamental Research Funds for the Central Universities. Z. Shen acknowledges support from the Natural Science Foundation of
Beijing, China (grant no. 1252020) and NSFC (grant no. 12572304). Y. Xu acknowledges support from NSFC (grant no. 12072010) and the Fundamental Research Funds for the Central Universities (GW2025-ZY-04).}
\backsection[Declaration of interests]{The authors report no conflict of interest.}
\backsection[Data availability statement]{Data supporting this work are openly available at \url{https://github.com/shiyuanhu/microroller_confinement}.}
\backsection[Author contributions]{H. Gao and N. Xie contribute equally to this work.}
\backsection[Author ORCIDs]{H. Gao, https://orcid.org/0000-0002-6370-9709; N. Xie, https://orcid.org/0009-0008-6573-5485; Z. Shen, https://orcid.org/0000-0003-4544-6463; X. Hu, https://orcid.org/0009-0002-2474-1287; S. Hu, https://orcid.org/0000-0002-8415-4263; Y. Xu, https://orcid.org/0000-0003-4322-244X.}

\appendix
\section{Asymptotic solutions of the translational velocity}\label{sec:AppendixA}
To validate our numerical method, we consider a spherical particle rotating near an infinite planar plate with angular velocity $\boldsymbol{\Omega} = \Omega\hat{\boldsymbol{y}}$. The resulting translational velocity $\boldsymbol{U}_0 = U_0\hat{\boldsymbol{x}}$ can be obtained using the Lorentz reciprocal theorem~\citep{Masoud2019} and asymptotic solutions of Stokes equations~\citep{Goldman1967}. Denote the flow field of the main problem as $\boldsymbol{u}(\boldsymbol{x})$ and the stress field as $\mathsfbi{\sigma}(\boldsymbol{x})$. We first introduce an auxiliary problem of a sphere undergoing pure translation with a given velocity $\boldsymbol{U}' = U'\hat{\boldsymbol{x}}$ parallel to the plate. The corresponding velocity and stress fields are $\boldsymbol{u}'(\boldsymbol{x})$ and $\mathsfbi{\sigma}'(\boldsymbol{x})$. The Lorentz reciprocal theorem states that
\begin{equation}\label{lorentz1}
\int_{S_{\mathrm{p}} + S_{\mathrm{w}} + S_\infty} \boldsymbol{n}\cdot \mathsfbi{\sigma}'\cdot \boldsymbol{u} \,\mathrm{d}\boldsymbol{x} = \int_{S_{\mathrm{p}} + S_{\mathrm{w}} + S_\infty} \boldsymbol{n}\cdot \mathsfbi{\sigma}\cdot \boldsymbol{u}' \,\mathrm{d}\boldsymbol{x},
\end{equation}
where $S_\infty$ is a bounding surface at infinity and $\boldsymbol{n}$ is the outward normal of the fluid domain. Applying the boundary conditions and the force-free condition for the main problem, $\int_{S_\mathrm{p}} \boldsymbol{n} \cdot \mathsfbi{\sigma}\, \mathrm{d} \boldsymbol{x} = 0$,
we obtain
\begin{equation}\label{reduced_lorentz1}
U_0 F'(U') = \Omega \int_{S_{\mathrm{p}}} \boldsymbol{n}\cdot \mathsfbi{\sigma}' \cdot \left[
\hat{\boldsymbol{y}} \times (\boldsymbol{x}-\boldsymbol{r}_\mathrm{c})\right]\,
\mathrm{d}\boldsymbol{x},
\end{equation}
where we have used $\boldsymbol{u}(\boldsymbol{x}) = U_0\hat{\boldsymbol{x}} + \Omega \hat{\boldsymbol{y}} \times (\boldsymbol{x} - \boldsymbol{r}_c)$ for $\boldsymbol{x} \in S_\mathrm{p}$, and $F'(U')$ is the drag force required to translate the particle with velocity $U'\hat{\boldsymbol{x}}$. In (\ref{reduced_lorentz1}), the translational velocity $U_0$ is unknown, and the rotational velocity $\Omega$ determines the driving condition. We introduce another auxiliary problem, namely, a particle held fixed in space while rotating with $\boldsymbol{\Omega} = \Omega\hat{\boldsymbol{y}}$ parallel to the plate. The solution of the second auxiliary problem, $\tilde{\boldsymbol{u}}(\boldsymbol{x})$ and $\tilde{\mathsfbi{\sigma}}(\boldsymbol{x})$, is related to that of the first auxiliary problem through
\begin{equation}\label{lorentz2}
\int_{S_{\mathrm{p}} + S_{\mathrm{w}} + S_\infty} \boldsymbol{n}\cdot \tilde{\mathsfbi{\sigma}}\cdot \boldsymbol{u}' \,\mathrm{d} \boldsymbol{x} = \int_{S_{\mathrm{p}} + S_{\mathrm{w}} + S_\infty} \boldsymbol{n}\cdot \mathsfbi{\sigma}' \cdot \tilde{\boldsymbol{u}} \,\mathrm{d} \boldsymbol{x}.
\end{equation}
Applying the boundary conditions, we obtain
\begin{equation}\label{reduced_lorentz2}
- U' \tilde{F}(\Omega) = \Omega \int_{S_{\mathrm{p}}} \boldsymbol{n}\cdot \mathsfbi{\sigma}' \cdot \left[
\hat{\boldsymbol{y}} \times (\boldsymbol{x}-\boldsymbol{r}_\mathrm{c})\right]\,
\mathrm{d} \boldsymbol{x},
\end{equation}
where $\tilde{F}(\Omega)$ is the drag force on the particle due to a pure rotational motion. Combining (\ref{reduced_lorentz1}) and (\ref{reduced_lorentz2}), the translational velocity in the main problem can be written as
\begin{equation}
U_0 = -\frac{U' \tilde{F}(\Omega)}{F' (U')}.
\end{equation}
Using the asymptotic expressions of $\tilde{F}(\Omega)$ and $F'(U')$ in the two limits $h/a \ll 1$ and $h/a \gg 1$~\citep{Goldman1967}, we obtain that, for $h/a \ll 1$,
\begin{equation}\label{U_asympt1}
U_0 = \Omega a \dfrac{\dfrac{2}{15}\ln\left(\dfrac{h}{a}\right) + 0.2526}{-\dfrac{8}{15}\ln\left(\dfrac{h}{a}\right) + 0.9588},  
\end{equation}
and for $h/a \gg 1$
\begin{equation}\label{U_asympt2}
U_0 = \Omega a \left[\frac{1}{8}\left(\frac{a}{z_\mathrm{c}}\right)^4\left(1-\frac{3}{8}\frac{a}{z_\mathrm{c}}\right)\right] \left[1-\frac{9}{16}\left(\frac{a}{z_\mathrm{c}}\right) + \frac{1}{8}\left(\frac{a}{z_\mathrm{c}}\right)^3 - \frac{45}{256}\left(\frac{a}{z_\mathrm{c}}\right)^4 - \frac{1}{16}\left(\frac{a}{z_\mathrm{c}}\right)^5\right],
\end{equation}
where the height of the particle center is $z_\mathrm{c} = h+a$.

\bibliographystyle{jfm}
\bibliography{jfm}

\end{document}